# Probability, Logic and Objectivity
*The concept of probability of Carl Stumpf*

## Contents

## Abstract

We will show that Carl Stumpf's interpretation of the concept of probability is best understood as that of an objective Bayesian. First we analyse Stumpf's work in relation to that of his contemporary Johannes von Kries, and after that we will discuss various ways in which Stumpf's probability-concept has been construed. By showing that the construals of Stumpf's account by Hans Reichenbach, Richard von Mises and Andreas Kamlah are unfair – and at some points incorrect – we uncover the aspects that are essential to Stumpf's probability interpretation.

## 1. Introduction

It has been claimed that some time before the turn of the 19$^{th}$ into the 20$^{th}$ century there was a paradigm-shift *à la* Kuhn in our scientific views on probability-theory[1]. The validity of this claim (or at least its content) is not generally agreed upon[2], but it is not contested that this was a turbulent period in the history of probability theory. The existence of such a turbulent period is supported by the fact that the issues that were debated and the arguments that were used in discussions about probability theory in this period are ubiquitous in 20$^{th}$ century philosophy of science. Obvious traces of the argumentations of John Stuart Mill and John Venn can be found in every frequentist probability interpretation almost 150 years after these scholars wrote their works.

In this paper we focus on a specific part of the probability-debate that took place in the generation after that of Venn and Mill. In the 1880's and 1890's Johannes von Kries and Carl Stumpf both defended their own reading of the 'classical' definition of probability that was due to Laplace. Just as other parts of the probabilistic debate the traces of the debate between Stumpf and von Kries extend very far. We will analyse several representations of Stumpf's side of the debate, and we will argue that in none of these representations Stumpf's view is accurately classified.

---

[1] (Kruger, Daston, & Heidelberger, 1987)
[2] (Hacking, Was there a probabilistic revolution, 1800-1930?, 1983)

In the next section we will introduce Laplace's definition of probability to prepare for an analysis of the arguments of Stumpf and von Kries themselves. The analysis of these arguments are the subject of the third and fourth sections. In the fifth section we will describe several ways in which Stumpf's view on probability has been understood. This description will be followed by a detailed investigation of the modern interpretations of probability which are closest to the view of Stumpf. In the final section of this paper we apply the results of our investigation to the various representations of Stumpf and draw our own conclusion as to how best to classify Stumpf's probability-concept.

## 2. Interpretations of Laplace

The 'traditional' interpretation of probability is that defended by Laplace in his *Philosophical essay on probabilities* (Laplace, 1812). In this work, Laplace considers the concept of probability to be applicable in the following way:

> *"The theory of chance consists in reducing all events of the same kind to a certain number of cases equally possible, that is to say, to such as we may be equally undecided about in regard to their existence, and in determining the number of cases favourable to the event whose probability is sought. The ratio of this number to that of all the cases possible is the measure of this probability, which is thus simply a fraction whose numerator is the number of favourable cases and whose denominator is the number of all cases possible."* (Laplace, 1812, p. 6)

The apparent straightforwardness of this quotation hides a number of difficulties. For instance, it is not clear from Laplace's words which 'events' should be regarded as 'of the same kind'. It is therefore also not clear to which events the probability-concept should be applied. The related – but not identical – difficulty we shall be concerned with here is that of the cases which are 'equally possible'. How can we justify the statement that certain cases or events are equally possible?

The concept of equipossibility had been around for at least a century before Laplace wrote his *essay*. There are two problems with the concept of equipossibility and its use in interpreting probability. The first problem is that we must discover what are the cases that are said to be equipossible in Laplace's definition. The second problem is that – once it is clear what they are – we must somehow justify the inference from equipossibility to equiprobability (without such an inference, Laplace's definition would not be about probability). It is not clear how Laplace can distinguish equipossibility from equiprobability and it is obvious that the use of the concept of equiprobability in defining probability is problematic – to say the least (Hacking, 1971). This problem has been addressed by many later authors (such as (von Kries, 1886), (von Mises, 1928), and (Reichenbach, 1949)). However, because the first problem of equipossibility plays a more important role in the work of Stumpf, we shall not consider the justification of the inference from equipossibility to equiprobability and focus on the problem of reference – what is the nature of the favourable and the possible cases in Laplace's definition of probability?

Laplace says in his *essay* that probabilistic judgments are partly relative to our knowledge and partly relative to the limits of this knowledge. When we want to describe some chance-event about whose outcome we are uncertain we use our knowledge of the situation to determine the numbers of favourable and possible cases. Furthermore, it is necessary that we have no knowledge whatsoever to prefer any of the cases over the others: we must be "*equally undecided*" about them. This means that probabilistic judgments about an event say something about the level of our knowledge about this event. Laplace is not always clear about this[3]. At certain points in his *essay* he states that the equipossibility must always correspond to some physical equality. This would lead us to believe that probabilistic judgments say something about objective, physical reality (as opposed to our incomplete knowledge thereof). We may conclude that Laplace is ambiguous about this matter; strictly speaking, it does not follow from the classical probability interpretation that the favourable and the possible cases in the probability definition should refer to either our (deficient) knowledge or to some physically existing structure.

## 3. Johannes von Kries

A fervent defender of an interpretation of probability in terms of objectively existing structure is the German physiologist Johannes von Kries. In 1886 von Kries wrote an account of his interpretation in his `Principien der Wahrscheinlichkeitsrechnung – eine logische Untersuchung' (von Kries, 1886). In this work von Kries invokes the notion of *event spaces* and a rule to assign measures to them.

Consider the example of the roll of a die. Together the roll's outcomes constitute a `universe of events', in which each of the six possible outcomes has its own `event-space'. But these event-spaces can be split up further: each outcome can be realized by very many microscopic configurations that all manifest themselves macroscopically in the same way (which is why they belong to the same event-space). The probability of an outcome can therefore be calculated if we are able to somehow assign a measure to this `space' of more fundamental microscopic events. This thought leads von Kries to define the probability of a specific outcome as the ratio of the size of its specific event space to the size of the totality of all event-spaces, corresponding to all possible outcomes. It is then possible to say which events are equiprobable – namely the events whose event spaces are of equal size. Given the symmetries of the situation, this procedure plausibly leads to assigning equal probabilities to all possible results of the throw of a fair die: one in six.

But what if our die is biased, loaded in such a way that the sides are not equipossible? Von Kries argues that in such cases we could and should trace back the causal nexus until we find event-spaces that do not split up anymore if the causal nexus is traced back even further. Suppose we do so in the case of the die: we start tracing back the chain of causes, going to a more fundamental level of description. Given the asymmetries in the physical situation (the die is loaded) we shall find that the number of initial states, in a (sub)microscopic description, that leads to one outcome will not be the same for all outcomes (we assume that the mechanism of casting the die is the same in all cases). When we go back even further in the causal nexus, allowing even more basic descriptions, these initial microstates will not split up any further, at least not in an asymmetric way. Von Kries dubs the event-spaces that do not split up further `elementary' event-spaces. The basic principle of von Kries' method is to take these elementary event-spaces as equipossible. So we have to continue our causal analysis until we find `simple, non-composite' causes; the elementary event spaces we thus ultimately arrive at will perhaps be defined in terms of the states of the individual atoms of which the die consists[4]. So for every chance event there are elementary event-spaces, but not every possible outcome needs to have the same number of elementary event-spaces associated with it. If, in the case of the biased die, we ask for the chance of six coming out, we should consider the totality of possible microscopic configurations (N), and determine the number of these that will lead to the desired outcome (n). The sought-for probability is then the ratio n/N.

---

[3] Cf. (Hacking, Was there a probabilistic revolution, 1800-1930?, 1983, p. 353) for a similar conclusion about Laplace.
[4] The atomic concept had not yet gained general acceptance in von Kries' time, but 'small quantities of matter' could serve a similar explanatory role.

# 4. Carl Stumpf

Seven years after von Kries had published his defence of an objectively based theory of probability which featured his notion of event-spaces, the psychologist and philosopher Carl Stumpf (1848-1936) held a lecture for the Bavarian Academy of the sciences (*On the concept of mathematical probability* (Stumpf, 1892a) in which he criticised von Kries' objective interpretation of probability. In his magnum opus `Erkenntnislehre', published three years after his death (Stumpf, 1940), Stumpf even writes that it was von Kries' theory of event-spaces that had motivated him to put into words his own views on the matter. In an addendum to the lecture held in 1892 (*On the application of the mathematical concept of probability to a part of a continuum*) (Stumpf, 1892b) Stumpf says that it does not follow from Laplace's theory that the equipossibility of cases should be founded on physical symmetries[5]. What Stumpf does agree on is Laplace's general definition of probability:

> *"We say that a certain event has a probability of n/N if we can regard it as one of n favourable cases within a total of N possible cases, of which we know only one is real, but we don't know which."*[6]

Stumpf argues that there is a requirement that this definition must meet. Probability must be defined in such a way that statements about degrees of probability of the occurrence of some event correspond to the degrees of rational expectation ("vernünftigen Erwartung" (Stumpf, 1892a, p. 56)) about this event that any rational observer would actually have. Stumpf says that "probability-theory is nothing but the mathematical justification of common sense." (Stumpf, 1892a, p. 39). Stumpf recognises as the central feature of the concept of probability that probabilistic statements are disjunctive statements. The different members of such a probabilistic disjunction correspond to the favourable and the possible cases of Laplace. The foremost requirement for such a probabilistic disjunction is that all available knowledge about a physical situation is used when formulating a probabilistic statement about this situation. A demand that springs from this requirement is that also all causal knowledge should be used.

We will explain this with an example of a toss with a symmetrical coin. The coin has two identical sides, so the probabilistic disjunction consists of two members. Suppose, however, that we know that in this particular experiment the mechanical causes are such that the two possible outcomes correspond to two sets of initial conditions for the toss whose cardinality is not equal. In that case, Stumpf argues, we should use that causal knowledge to arrive at a disjunction which is different from the original disjunction that was based on knowledge of the coin only. Stumpf does not argue that all probabilistic knowledge should ultimately be based on causal knowledge. However, if such knowledge is available then the members of the disjunction should be traced back to their causes and the members of the updated probabilistic disjunction should describe these causes.

---

[5] „In meinem Vortrage, ‚Über den Begriff...‘ habe ich in Consequenz der Wahrscheinlichkeitsdefinition von Laplace gegenüber neueren Auffassungen daran festgehalten, dass zur Wahrscheinlichkeitsbestimmung physische Gleichheit der sog. gleichmöglichen fälle nicht erforderlich sei." (Stumpf, 1892b, p. 681)
[6] „Jede beliebige Urteilsmaterie nennen wir n/N Wahrscheinlich, wen wir sie auffassen können als eines von n gliedern (günstigen Fällen) innerhalb einer gesamtzahl von N gliedern (möglichen Fällen), von denen wir wissen, dass eines und nur eines wahr ist, dagegen slechterdings nicht wissen welches." (Stumpf, 1892a, p. 48)

Stumpf's interpretation of probability differs markedly from the interpretation espoused by von Kries. It is true that Stumpf, just as von Kries, believes that we should use available causal knowledge in formulating statements of probability, but Stumpf does not agree with von Kries that it follows from that that probabilistic judgments should be based solely on physically existing structures. Stumpf's side in the debate with von Kries on probability has been represented in various ways. In the following section we will discuss several of such representations. Their analysis will provide us with what is necessary to reach our own conclusion on how Stumpf's work should be interpreted and allow us to accurately place Stumpf within the history of probability-theory.

## 5. Interpretations of Stumpf

The first representation of Stumpf's views on probability which we will be concerned with here is that by the philosopher of science Hans Reichenbach. In 1915 Reichenbach wrote his dissertation in which he formulated a neokantian interpretation of the concept of probability (The Concept of Probability in the Mathematical Representation of Reality, 1915)[7]. Reichenbach begins his work with an account of what he believes should be the major point of concern in the contemporary debate about probability. In order to prepare for a discussion of his own view Reichenbach represents the interpretations of Stumpf and von Kries as two opposing ways of interpreting Laplace's classical definition. Reichenbach argues that if the concept of probability is to play a role in science, then it should be wholly objectively interpreted – a change in probability can only exist if there is a corresponding change in physical reality. Without treating Stumpf's views in any detail Reichenbach virtually dismisses Stumpf's view as naively subjective. Reichenbach does not doubt the internal consistency of Stumpf's probability concept, but he argues that if we adopt Stumpf's view, we are not justified in taking probability judgments as a basis for rational expectation:

> "Der Stumpfsche Wahrscheinlichkeitsbegriff ist zwar in sich widerspruchslos, und kann deshalb nicht falsch genannt werden, da Definitionen willkürlich sind. Aber er leistet jedenfalls nicht das, was man allgemein von einem Wahrscheinlichkeitsbegriff verlangt. Denn er ist nicht geeignet, ein Mass der vernünftigen Erwartung abzugeben." (Reichenbach, 1915, p. 46)

Reichenbach was not alone in his appraisal of Stumpf's views on probability. One of the founders of the 20th century frequentist interpretation of probability, the Ukrainian born mathematician Richard von Mises, calls Stumpf the main representative (*Hauptvertreter*) of the subjectivist interpretation of probability (von Mises, 1928, p. 269). A more recent (and more elaborate) account of Stumpf's probability concept is due to Andreas Kamlah (The decline of the Laplacian theory of probability: a study of Stumpf, von Kries, and Meinong, 1987). Kamlah argues that Stumpf's interpretation should be understood as equivalent with Carnap's *logical interpretation*. In the course of an analysis of Stumpf's view, Kamlah provides us with a definition of Stumpf's probability that is "better than his [Stumpf's] own" (Kamlah, 1987, p. 102). Kamlah gives a formalised version of Stumpf's own definition in terms of propositions which is in line with Stumpf's remark that statements of probability are statements about logical relations ("*logische Zusammenhänge*") (Stumpf, 1892a, p. 99). It should be clear here that Kamlah attempts to precisify Stumpf's definition in order to explicate what Stumpf actually meant (and not in order to give his own view on the matter). In the next section we will analyse several different contemporary interpretations of the probability-concept that are intimately related to the views of Stumpf.

---

[7] Cf. (Eberhardt, 2011) for an appraisal of Reichenbach's dissertation and (Benedictus & Dieks, 2014) for a brief exposition of Reichenbach's rejection of Stumpf.

# 6. Bayesianism

In the fourth section of his main lecture in 1892 (1892a), Stumpf confesses himself to be a *subjective Bayesian* (Stumpf, 1892a, p. 96). It will become apparent that this remark, although it appears only in the final section of Stumpf's account, is very important for the issues dealt with in this paper.

The central idea of the Bayesian interpretation of probability is that degrees of probability correspond to the degrees of belief of some observer. For example, consider the probability of some event taking place and assume that there is a certain piece of evidence in favour of the event's actual happening (in case of a die-throw the evidence might be an observed relative frequency within a sequence of outcomes of earlier throws). The Bayesian now distinguishes between prior and posterior probability. The prior probability corresponds to the observer's degree of belief in the actual happening of the event before the evidence comes into play, and the posterior probability corresponds to the observer's degree of belief in the actual happening of the event after the evidence has come into play. Let **A** be the event whose probability we are interested in, and let **B** represent the evidence in **A**'s favour. If we denote the prior probability of **A** as **P(A)** and the posterior probability of **A** as **P(A|B)** then the Bayesian holds that the change in degree of belief should follow Bayes' formula (otherwise the observer would not be rational):

$$P(A \mid B) = \frac{P(B \mid A) \cdot P(A)}{P(B)} \qquad (1)$$

It is not the validity of this formula that distinguishes Bayesianism from other interpretations of probability. Rather, this distinction lies in the fact that the Bayesian equates the prior and posterior degrees of probability that are fed into the above equation as degrees of belief.

Two types of Bayesianism are current in the literature[8]: objective and subjective Bayesianism. The distinction between the two types is clear, but it is not always unequivocal to which of the camps a Bayesian mathematician or philosopher belongs. This ambiguity results from the fact that not many Bayesians would call their view either purely subjective or purely objective[9]. Although the different types (together with most other interpretations of probability) agree in that any rational observer should update her degrees of belief following equation (1), the objective and the subjective Bayesian differ in their characterisation of prior probabilities.

In a purely objective Bayesian view there are rational constraints that fully determine the prior probabilities that go into equation (1). If certain evidence regarding some event is available to an observer (ie. if the body of knowledge an agent possesses is fixed), then according to a purely objective Bayesian there is only one degree of belief regarding the probability of the event available to this observer. This uniquely determined degree of belief is what a purely objective Bayesian calls the degree of probability.

---

[8] Cf. (Talbott, 2008) for a concise account of how subjective Bayesianism differs from objective Bayesianism.

[9] There certainly are some of such 'extremists'. Bruno de Finetti is an example of a purely subjective Bayesian. (Talbott, 2008) maintains that no contemporary Bayesian is purely objective. Also, (Chalmers, 1976, p. 178) argues that purely objective Bayesianism is not a feasible position as its adherents cannot cope with empirically equivalent hypotheses.

In a purely subjective Bayesian view such rational constraints on prior probabilities are wholly absent. According to a purely subjective Bayesian the prior probabilities can take on any value between zero and one; for an observer to be rational it is enough that she updates her degrees of belief following Bayes' formula (1). Most versions of Bayesianism are somewhere between purely objective and purely subjective;[10] they vary in their appraisal of the extent to which there are rational restraints on the Bayesian prior probabilities.

The *logical interpretation* of probability is intimately related to Bayesianism.[11] The modern-day form of the logical interpretation is mostly due to the work of Rudolf Carnap. Carnap has written about his logicality view on probability in many earlier publications, but the first book in which he gives a detailed account of his probabilistic logic appeared in 1950 (Logical Foundations of Probability). In this work Carnap defines probability as a relation between propositions within a language. To show how that works he considers a simple model of a language which consists solely of names and predicates. Such a language can be used to describe a model-world which consists only of a number of objects to which the names refer. In our model-world there is only one characteristic which each of the objects either has or does not have. Provided with such a language we can draw up a full description of our world. Such a description – which lists all objects whether they have the characteristic or not – Carnap calls a *state description*. Different state descriptions correspond to different linguistic statements about our model-world.

Probability, in Carnap's logical view, is *defined* as a measure **m(-)** over all possible state descriptions. This measure automatically extends over all possible sentences within the language that we use, so Carnap's probability could be said to express nothing but a logical relation between sentences within a language. The probability measure in turn allows Carnap to define a *confirmation function*, **c(h,e)**, which formalises the relation between propositions about evidence (**e**) and propositions which express a hypothesis (**h**):

$$c(h,e) = \frac{m(h \& e)}{m(e)} \qquad (2)$$

The details of how **c** and **m** can be defined in terms of propositions about our model-world (and how **h** and **e** should be construed) need not concern us here. What is important is that we realise that in Carnap's logical interpretation degrees of probability correspond to degrees of confirmation of some hypothesis, and that this manifests itself as logical relations between propositions within a language.

As the similarity between equations (1) and (2) already suggests the logical interpretation is closely related to both types of Bayesianism. In modern literature the logical interpretation is often called a 'limiting case' of objective Bayesianism[12]. To see why that is, we return to our example of a situation in which an observer is confronted with a certain body of evidence. For someone who interprets probability in a logicist sense, the only guide on the road from evidence to judgments of probability is logical deduction. About the purely objective Bayesian we know that she will say that the prior degrees of belief of the observer are completely determined by the evidence, and that the posterior degrees of belief follow from the priors with logical rigour. A purely objective Bayesian is therefore fully rational. It is clear from this in what sense the logical interpretation can be regarded as a limiting case of (objective) Bayesianism. The sense is one of the 'degree of rationality'.

---

[10] E.g. the Bayesian views of E.T. Jaynes and J.M. Keynes.
[11] For the contents of this paragraph we lean heavily upon (Hájek, 2011).
[12] Cf. (Talbott, 2008) and (Williamson, 2009).

But there is also an essential difference between Carnap's logical interpretation and that of the objective Bayesian. The crux is that in the logical interpretation statements of probability express relations between propositions within a particular (formal) language, whereas for the objective Bayesian language plays no role[13]. The fact that in the logical interpretation statements of probability express linguistic relations makes the validity of a probability statement depend on the particular language[14] in which the statement is made. In a given situation more than one statement of probability can be considered 'correct' as a result of this. In fact, there is always a 'continuum' of possible languages to describe a chance-event, each corresponding to a different degree of probability for the event. For the objective Bayesian, on the other hand, the situation is less liberal. The objective Bayesian holds that degrees of probability are degrees of belief. We may assume that a rational agent has only one particular degree of belief (schizophrenia aside). It follows that in a given situation there can be only one statement of probability that is the correct one – that is the one that corresponds to the degree of belief of the rational agent. It has become clear that there is a crucial difference between the logical interpretation of probability and that of the objective Bayesian.

## 7. Conclusion

The preceding sections of this paper have put us in a position where we can compare the different perspectives on Stumpf's concept of probability and properly classify Stumpf's interpretation within the history of probability-theory.

The first account of Stumpf's probability concept that we discussed was that of Reichenbach. Although Reichenbach judged Stumpf's account to be coherent, he also believed that the element of subjectivity that is inherent to Stumpf's view renders Stumpf's probability concept unusable for science. The arguments of Reichenbach evidently depend on what notion of 'objectivity' we adopt, and, consequently, how we characterise subjectivity. Reichenbach argues that a change in probability must correspond to a physical change. That is how Reichenbach characterises objectivity within the context of probability. According to Stumpf, on the other hand, the existence of an underlying physical change which is responsible for a change in probability is merely a hypothesis. Stumpf's idea about the objectivity of probability derives from his definition (which we encountered on p.4). For Stumpf the concept of probability is objective in very restricted sense. Stumpf's notion of objectivity follows from the requirement that degrees of probability conform to the degrees of expectation of a rational observer.

> *„Nennen wir nun ‚objektiv gültig' dasjenige, was von alle Subjecten bei gleicher Urteilsmaterie anerkannt werden muss, so ist das Wahrscheinliche objectiv gültig"* (Stumpf, 1892a, p. 55).

---

[13] One might counter that *any* statement that anyone makes is a statement in a certain language. It follows that language always plays a role (if it ever does). Then how can it be that language plays no role for the objective Bayesian?

For the objective Bayesian a degree of probability corresponds to a degree of belief, but that does not mean that every degree of belief uniquely corresponds to a probability statement. The language used by the logicist is a formal language, whereas that of the objective Bayesian may be a natural language.

[14] Provided that this language is a *formal* language, the choice of a particular language corresponds to the definition of a certain measure.

A correct understanding of this notion of objectivity is enough to clear Stumpf's name in the face of Reichenbach's accusations. Reichenbach's claim that Stumpf's probability concept yields no basis for rational expectation is clearly based upon a careless reading of Stumpf's account. Reichenbach's claim becomes even more surprising if we read the following words of Stumpf:

> „Die mathematische Wahrscheinlichkeit wird auch als *Mass unserer vernünftigen Erwartung* bezeichnet." (Stumpf, 1892a, p. 56) (Italics in the original)

Reichenbach clearly misjudged Stumpf. There is more merit to von Mises' classification of Stumpf as a subjectivist, because the knowledge of a subject plays an essential role in Stumpf's account. We have seen, however, that Stumpf's probability-concept is subjective only in a very restricted sense of the word. Kamlah's characterisation of Stumpf's probability concept as a logical concept of probability is more apropos than the characterisations by Reichenbach and von Mises. Kamlah correctly states that Stumpf's probability concept is of a logical nature: Stumpf literally states that statements of probability express logical relations between propositions. However, the logical interpretation does not incorporate Stumpf's notion of objectivity. We have seen that in the logical interpretation of Carnap (to whom Kamlah explicitly refers (Kamlah, 1987, p. 108)) an infinite number of different probability judgments may apply to a fixed body of evidence (due to the freedom in the choice of the probability measure (**m**) used). In Stumpf's account such freedom is absent. We see that Kamlah, just as Reichenbach and von Mises, has misconstrued Stumpf's probability concept.

Although it fails to accurately incorporate Stumpf's notion of objectivity Kamlah's logical understanding of Stumpf is more to the point than Reichenbach's accusation of unbridled subjectivism. Stumpf himself claims that his interpretation of probability is that of the subjective Bayesian. However, this claim is inconsistent with his own notion of objectivity. We have seen that the subjective Bayesian believes that posterior probabilities are logically connected to prior probabilities, but that the choice of these prior probabilities themselves is free (and not necessarily rational). Such a free choice squarely contradicts Stumpf's statement that probability has an objective nature. *We therefore conclude that Stumpf's probability-concept is best understood as that of the objective Bayesian*.

We end our paper with a speculative remark. To understand why Stumpf adopted the stance of the objective Bayesian as opposed to von Kries' physical event spaces, it is instrumental to be conscious of the fact that he did so as a psychologist. In the introductory section of his main lecture in 1892 Stumpf states that interpreting the theory of probability is of great importance for science. Not only has the role of probability theory in the exact sciences grown, but it is also of fundamental importance to the 'moral sciences' and the humanities (*Moral- und Geisteswissenschaften*) (Stumpf, 1892a, p. 37). Stumpf's reason for believing this derives from his view that all science ultimately depends on statistics. In psychological research it is not always easy, or perhaps sometimes even impossible, to measure the event spaces that are relevant to von Kries' probability concept. Therefore, the adoption of von Kries' account of probability would rule out (large parts of) psychology as a science. Of course Stumpf could not accept a concept of probability that would undermine his own work as a scientist. We claim that Stumpf embraced objective bayesianism as a way of legitimising his view of psychology as a science.

# Acknowledgments

We would like to thank Professor Dennis Dieks for providing innumerous helpful suggestions and many valuable comments.